\documentclass[pra,twocolumn,showkeys,showpacs,superscriptaddress,floatfix,longbibliography]{revtex4-1}

\usepackage[T1]{fontenc}
\usepackage[latin9]{inputenc}
\usepackage{epsfig,psfrag}
\usepackage{latexsym}
\usepackage{graphicx}
\usepackage{dcolumn}
\usepackage{amssymb}
\usepackage{amsmath}
\usepackage{bm}
\usepackage{mathrsfs} 
\usepackage{bigints}
\usepackage{upgreek}
\usepackage{color}
\usepackage{longtable}
\usepackage{xspace}
\usepackage[squaren]{SIunits} 
\usepackage{epstopdf}
\usepackage{float}
\usepackage[normalem]{ulem}
\usepackage{hyperref}
\usepackage{bbold}
\usepackage{bbm}

\definecolor{mygreen}{RGB}{20,148,20}

\usepackage{filemod}

\newcommand{\eq}[1]{Eq.~\eqref{eq:#1}}

\newcommand{\eqr}[2]{Eqs.~\eqref{eq:#1} to~\eqref{eq:#2}}

\newcommand{\ket}[1]{{\mid} {#1}{\rangle} }

\begin{document}

\title{Wave Packet Dynamics in Synthetic Non-Abelian Gauge Fields} 

\author{Mehedi Hasan}
\altaffiliation[Currently at ]{Cavendish Laboratory, University of Cambridge, Cambridge CB3 0HE, United Kingdom}
\affiliation{Nanyang Quantum Hub, School of Physical and Mathematical Sciences, Nanyang Technological University, 21 Nanyang Link, Singapore 637371, Singapore}
\affiliation{MajuLab, International Joint Research Unit IRL 3654, CNRS, Universit\'e C\^ote d'Azur, Sorbonne Universit\'e, National University of Singapore, Nanyang
Technological University, Singapore}
\author{Chetan Sriram Madasu}
\affiliation{Nanyang Quantum Hub, School of Physical and Mathematical Sciences, Nanyang Technological University, 21 Nanyang Link, Singapore 637371, Singapore}
\affiliation{MajuLab, International Joint Research Unit IRL 3654, CNRS, Universit\'e C\^ote d'Azur, Sorbonne Universit\'e, National University of Singapore, Nanyang
Technological University, Singapore}
\author{Ketan D. Rathod}
\altaffiliation[Currently at ]{Bennett University, Greater Noida 201310, India}
\affiliation{Centre for Quantum Technologies, National University of Singapore, 117543 Singapore, Singapore}
\affiliation{MajuLab, International Joint Research Unit IRL 3654, CNRS, Universit\'e C\^ote d'Azur, Sorbonne Universit\'e, National University of Singapore, Nanyang
Technological University, Singapore}
\author{Chang Chi Kwong}
\affiliation{Nanyang Quantum Hub, School of Physical and Mathematical Sciences, Nanyang Technological University, 21 Nanyang Link, Singapore 637371, Singapore}
\affiliation{MajuLab, International Joint Research Unit IRL 3654, CNRS, Universit\'e C\^ote d'Azur, Sorbonne Universit\'e, National University of Singapore, Nanyang
Technological University, Singapore}
\author{Christian Miniatura}
\affiliation{MajuLab, International Joint Research Unit IRL 3654, CNRS, Universit\'e C\^ote d'Azur, Sorbonne Universit\'e, National University of Singapore, Nanyang
Technological University, Singapore}
\affiliation{Centre for Quantum Technologies, National University of Singapore, 117543 Singapore, Singapore}
\affiliation{Department of Physics, National University of Singapore, 2 Science Drive 3, Singapore 117542, Singapore}
\affiliation{Nanyang Quantum Hub, School of Physical and Mathematical Sciences, Nanyang Technological University, 21 Nanyang Link, Singapore 637371, Singapore}
\affiliation{Universit\'e C\^ote d'Azur, CNRS, INPHYNI, Nice 06108, France}
\author{Fr\'ed\'eric Chevy}
\affiliation{Laboratoire de Physique de l'\'Ecole normale sup\'erieure, ENS, Universit\'e PSL, CNRS, Sorbonne Universit\'e, Universit\'e de Paris, F-75005 Paris, France}
\author{David Wilkowski}
\email{david.wilkowski@ntu.edu.sg}
\affiliation{Nanyang Quantum Hub, School of Physical and Mathematical Sciences, Nanyang Technological University, 21 Nanyang Link, Singapore 637371, Singapore}
\affiliation{MajuLab, International Joint Research Unit IRL 3654, CNRS, Universit\'e C\^ote d'Azur, Sorbonne Universit\'e, National University of Singapore, Nanyang
Technological University, Singapore}
\affiliation{Centre for Quantum Technologies, National University of Singapore, 117543 Singapore, Singapore}


\begin{abstract} 
It is generally admitted that in quantum mechanics, the electromagnetic potentials have physical interpretations otherwise absent in classical physics as illustrated by the Aharonov-Bohm effect. In 1984, Berry interpreted this effect as a geometrical phase factor. The same year, Wilczek and Zee generalized the concept of Berry phases to degenerate levels and showed that a non-Abelian gauge field arises in these systems. In sharp contrast with the Abelian case, spatially uniform non-Abelian gauge fields can induce particle noninertial motion. We explore this intriguing phenomenon with a degenerated Fermionic atomic gas subject to a two-dimensional synthetic $\mathrm{SU}(2)$ non-Abelian gauge field. We reveal the spin Hall nature of the noninertial dynamic as well as its anisotropy in amplitude and frequency due to the spin texture of the system. We finally draw the similarities and differences of the observed wave packet dynamic and the celebrated \textit{Zitterbewegung} effect of the relativistic Dirac equation.
\end{abstract}

\pacs{}

\keywords{}

\maketitle

Since the pioneering work of Lin and co-workers on synthetic magnetic fields \cite{lin2009synthetic}, intensive works on quantum simulators such as ultracold-gas platforms \cite{dalibard2011colloquium, atala2013direct, goldman2014light,zhai2015degenerate,zhang2019recent,cooper2019topological} or photonic circuits \cite{RevModPhys.91.015006} have been carried out to generate and explore artificial Abelian or non-Abelian gauge fields. The overarching objective is to explore geometrical and topological properties of quantum matter and materials. In particular, thanks to the noncommutative nature of the components of non-Abelian gauge fields, the eigenstates of the Hamiltonian are characterized by a momentum-dependent spin texture that leads to a myriad of phenomena such as spin phase separations \cite{lin2011spin,PhysRevA.94.061604}, topological Lifshitz transitions in a degenerate Fermi gas \cite{wang2012spin}, topological phases in a Bose-Einstein condensate \cite{putra2020spatial}, or the Josephson-like effect for interacting quantum gases \cite{hou2018momentum}. The band structure, the spin texture, and the topology properties of two-dimensional (2D) spin-orbit coupled ultracold-atom systems have been reported in \cite{wu2016realization,huang2016experimental,meng2016experimental}.

The coupling to a non-Abelian gauge field can take the form of a spin-orbit coupling (SOC) Hamiltonian \cite{dalibard2011colloquium},
\begin{equation} \label{eq:HSOC}
    \hat{H}_{\textrm{SOC}}=-\mathbf{\hat{p}}\cdot\mathbf{\hat{A}}/m,
\end{equation} 
where $\mathbf{\hat{p}}$ is the momentum operator of the particle, $m$ its mass, and $\mathbf{\hat{A}}$ the non-Abelian gauge field operator acting in the pseudospin space. For systems under $\mathrm{SU}(2)$ symmetry, the wave packet dynamics of a SOC system is predicted to exhibit an oscillatory behavior, similar to the \textit{Zitterbewegung} of the Dirac equation, i.e., a trembling motion of a particle associated with a quantum Rabi flopping of the pseudospin \cite{cserti2006unified,zhang2013zitterbewegung,vaishnav2008observing,merkl2008atomic,zhang2010driven,PhysRevLett.108.035302,argonov2016zitterbewegung,garreau2017simulating}. The oscillatory behavior of the wave packet has been experimentally studied in one-dimensional (1D) systems, where the SOC term reduces to a single component gauge field \cite{PhysRevLett.105.143902,gerritsma2010quantum,PhysRevA.88.021604,leblanc2013direct}. In this context, it has been shown that the \textit{Zitterbewegung} is present if the total Hamiltonian includes a scalar potential, which does not commute with the SOC Hamiltonian \cite{gerritsma2010quantum}. In the Dirac equation the scalar term is the mass operator of the particle-antiparticle system. In the 1D SOC Hamiltonians experimentally explored in photonic platforms \cite{PhysRevLett.105.143902}, trapped ions \cite{gerritsma2010quantum}, and ultracold gases \cite{PhysRevA.88.021604,leblanc2013direct}, the scalar potential is a Zeeman-like term.   

In this Letter, we report on studies of the atomic wave packet dynamics in a two-dimensional (2D) spatially uniform $\mathrm{SU}(2)$ non-Abelian gauge field. As a key feature, the wave packet shows oscillatory dynamics coming from the SOC Hamiltonian only, i.e., without any scalar potentials. The occurrence of the dynamics can be understood by deriving the time evolution of the velocity operator $\mathbf{\hat{v}}=(\mathbf{\hat{p}}-\mathbf{\hat{A}})/m$ in the Heisenberg picture. It leads to a noninertial force depending on the particle  momentum,
\begin{equation} \label{eq:Evol_V}
   m \frac{d{\bf \hat{v}}}{dt} = \frac{im}{\hbar} [\hat{H}_{\mathrm{SOC}},{\bf \hat{v}}] = -\frac{i}{m\hbar} \, {\bf \hat{p}} \times ({\bf \hat{A}} \times {\bf \hat{A}}).
\end{equation} 
Two important observations can be made from \eq{Evol_V}. First, even for an uniform gauge field, the 2D dynamics is strongly affected by the non-Abelian nature of the gauge field since the velocity is no longer constant when $({\bf \hat{A}} \times {\bf \hat{A}})$ is non zero. Second, since the velocity component along the momentum does not change in time, the nontrivial wave packet dynamics occurs in the plane transverse to the momentum. This locking of the dynamics at right angle of the momentum is reminiscent of the spin Hall effect \cite{beeler2013spin,kato2004observation}. Note that $\mathbf{\hat{p}}$ commutes with $\hat{H}_nontrivial\mathrm{SOC}$ for spatially uniform gauge fields and is then a constant of motion. In this case, the plane of oscillations does not change with time. Another striking feature that we will demonstrate later is the anisotropy of the wave packet dynamic in momentum space induced by the spintexture of the non-Abelian Hamiltonian.

\begin{figure}
\includegraphics[width=0.47\textwidth]{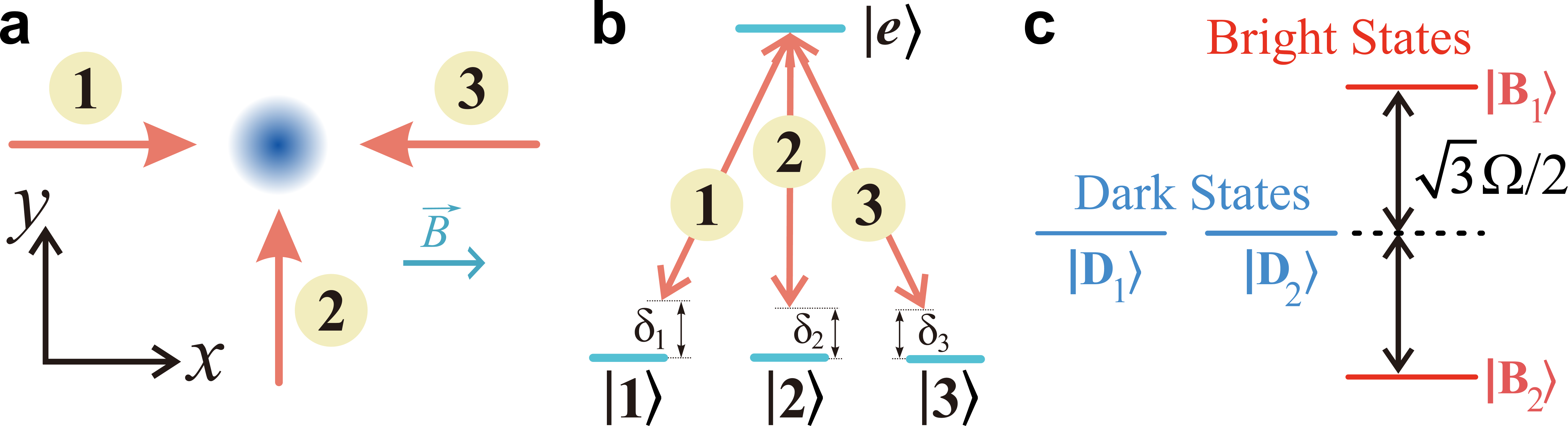}
\caption{{ (a)} Real-space configuration of the tripod laser beams. The three co-planar beams (red arrows) define the plane ($Ox,Oy$) orthogonal to the gravity pull. A $67$ G bias magnetic field, applied along $Ox$, allows us to isolate the tripod system among the $(^1S_0, F_g=9/2) \to (^3P_1, F_e=9/2)$ Zeeman manifold of the intercombination line. Counterpropagating beams 1 and 3 along $Ox$ have opposite circular polarizations and address $\sigma_+$ and $\sigma_-$ transitions, respectively. The orthogonal beam 2 is linearly polarized along $Ox$ and addresses a $\pi$ transition. 
{ (b)} The tripod laser beam $a$ drives the transition $\ket{a} \leftrightarrow \ket{e}$ with a detuning $\delta_a$ ($a=1,2,3$). The resonant Rabi frequencies are all equal to $\Omega/2\pi = 210$~kHz. 
{ (c)} The internal dressed-state basis of the system features two degenerate dark states (in blue) uncoupled to the laser beams and two bright states (in red) shifted from the dark states by $\pm\sqrt{3}\hbar\Omega/2$.} \label{fig:fig1a} 
\end{figure}

To generate our artificial non-Abelian gauge field, we use three quasiresonant, suitably polarized, laser beams; see Fig.~\ref{fig:fig1a}(a). These lasers operate on the $(^1S_0, F_g=9/2) \to (^3P_1, F_e=9/2)$ intercombination line at $689\,$nm (frequency linewidth $\Gamma/2\pi=7.5\,$kHz) of the fermionic strontium isotope $^{87}$Sr. They couple, in a tripod configuration, three Zeeman bare ground states $|a \rangle \equiv |F_g, m_F\rangle$, with $a=1,2,3$ and $m_F=5/2, 7/2, 9/2$ respectively, to the same excited state $\ket{e} \equiv \ket{F_e,m_F=7/2}$ and with equal Rabi frequencies $\Omega/2\pi = 210$~kHz, see Fig.~\ref{fig:fig1a}(b)~\cite{leroux2018non}. A bias magnetic field of $67$ G along the $x$ axis ensures that the states outside the tripod remain spectators (the Zeeman frequency shift of the excited state is around $7\,\mathrm{MHz}\gg\Omega/2\pi, \Gamma/2\pi$ \cite{PhysRevA.76.022510}). 

In the dressed-state picture, the internal Hamiltonian of the tripod system has two bright states, coupled to the laser fields, separated by $\pm\sqrt{3}\hbar\Omega/2$ from two degenerate zero-energy dark states, see Fig.~\ref{fig:fig1a}(c). The bright or dark state energy separation being large enough, the quantum state of the atoms remains and evolves in time in the dark state manifold \cite{leroux2018non}. The dark states representation in the bare-state basis reads
\begin{eqnarray}
    \ket{D_1}&=&\frac{1}{\sqrt{2}}\left(e^{-2ikx}\vert 1\rangle-e^{-ik(x+y)}\vert 2\rangle\right),\label{eq:D1}\\
	\ket{D_2}&=&\frac{1}{\sqrt{6}}\left(e^{-2ikx}\vert 1\rangle+e^{-ik(x+y)}\vert 2\rangle-2\vert 3\rangle\right), \label{eq:D2}
\end{eqnarray}
where $k$ is the tripod beams wave number. 

For resonant excitation ($\delta_a =0$), the Hamiltonian in the 2D dark-state manifold takes the form \cite{dalibard2011colloquium} 
\begin{equation} \label{eq:GFHamiltonianWithPhi} 
    \hat{H}_0 = \frac{\left(\mathbf{\hat{p}}-\mathbf{\hat{A}}\right)^2}{2m} + \hat{\Phi}.
\end{equation}
In the pseudospin representation of the dark-state manifold, the vector and scalar gauge field potentials ($\mathbf{\hat{A}}, \hat{\Phi}$) are represented by $2\times 2$ matrices with  entries $\mathbf{\hat{A}}_{jk}=i\hbar\langle D_j\vert\boldsymbol{\nabla}D_k\rangle$ and $\hat{\Phi}_{jk}=\left[\hbar^2\langle \boldsymbol{\nabla}D_j\vert \boldsymbol{\nabla}D_k\rangle-(\mathbf{\hat{A}}^2)_{jk}\right]/2m$ ($j,k = 1,2$) \cite{dalibard2011colloquium,SM}. 

For quasiresonant excitation ($|\delta_a|\ll\Omega$), the laser detuning contribution in the dark-state manifold reduces to an additional scalar matrix potential. We use it for two crucial purposes: to cancel the scalar term $\mathbf{\hat{A}}^2/2m+ \hat{\Phi}$ obtained by expanding the square in \eq{GFHamiltonianWithPhi}, and to perform a Galilean transformation into an inertial frame moving at an arbitrary velocity $-\mathbf{v}_0$ by adding a new term $-\mathbf{v}_0\cdot\mathbf{\hat{A}}$ (see Sec. C in Supplementary Material \cite{SM}). 
In this moving frame, and up to inessential constant terms proportional to unity, the Hamiltonian becomes 
\begin{equation} \label{eq:general_H}
    \hat{H} = \frac{\mathbf{\hat{q}}^2}{2m} - \frac{\mathbf{\hat{q}}\cdot\mathbf{\hat{A}}}{m},
\end{equation}
where $\mathbf{\hat{q}} = \mathbf{\hat{p}} + m\mathbf{v}_0$. We get the expected SOC Hamiltonian without scalar contribution to explore the 2D dynamical properties of this system. Note that the noninertial dynamic is not affected by the discarded spin-independent $\mathbf{\hat{q}}^2/(2m)$ term \cite{cserti2006unified}. 

\begin{figure}
\includegraphics[width=0.47\textwidth]{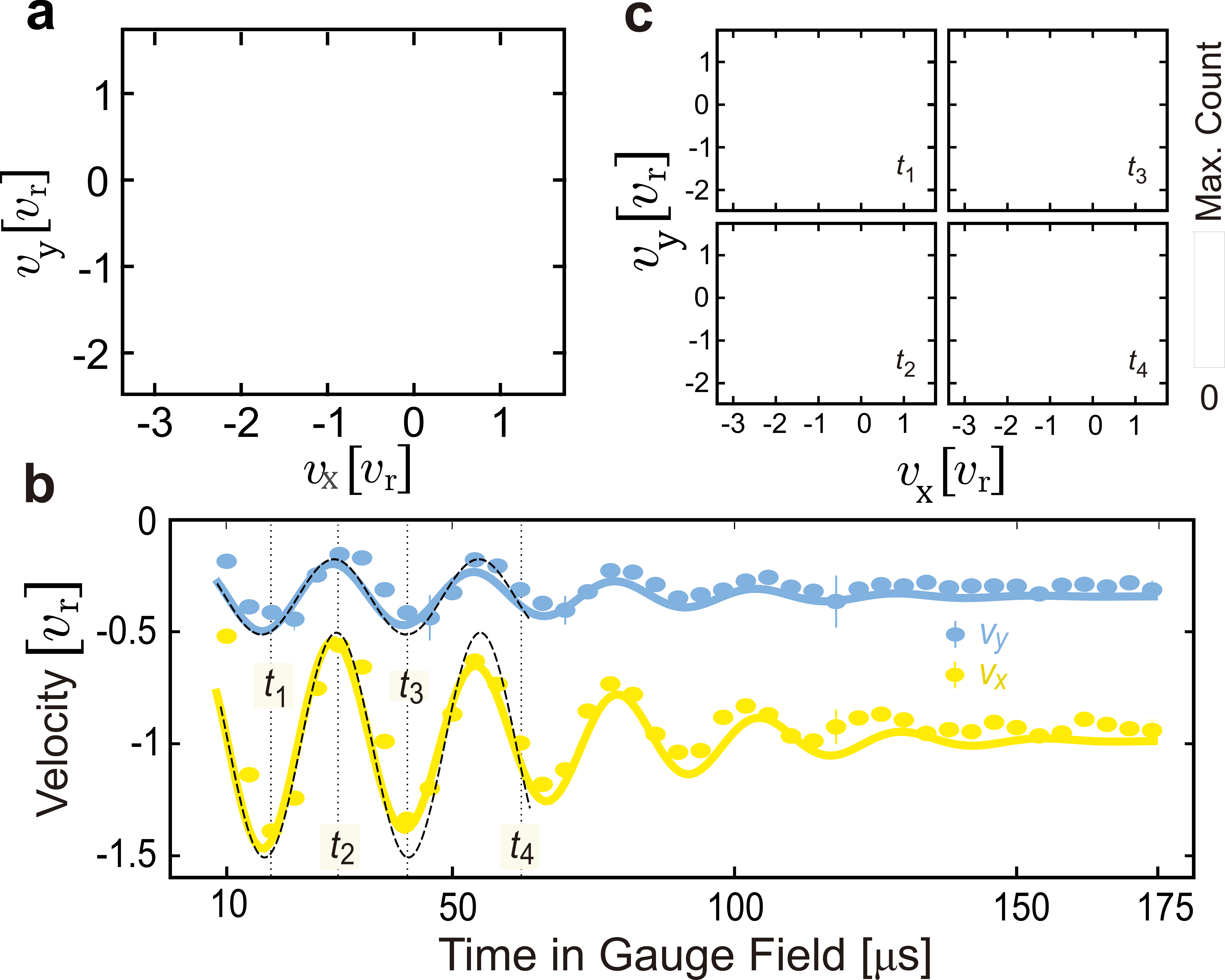}
\caption{{ (a)} Time-of-flight fluorescence image recorded at $v_0 =0$ after the system has been initialized in the dark state $\ket{D_2}$. The ballistic time is $9\,$ms. As expected from \eq{D2}, the measured velocity distribution shows three peaks centered at $\mathbf{v}=0$, $\mathbf{v}=-v_r(\hat{{\bf e}}_x+\hat{{\bf e}}_y)$ and $\mathbf{v}=-2v_r \hat{{\bf e}}_x$ ($v_r=\hbar k/m$ is the recoil velocity). These peaks correspond to states $\ket{3}$, $\ket{2}$, and $\ket{1}$, respectively. 
{(b)} Temporal oscillations of the Cartesian coordinates of the averaged velocity obtained for a boost velocity $(v_0=4\sqrt{2} v_r, \theta_0 = 0.6 \pi)$. The solid lines are obtained by numerically integrating the time evolution of the system in the gauge field, initialized in $\ket{3}$. We include the $10~\mu$s ramping stage of the tripod  beams and finite momentum distribution of our fermionic gas at $T=30(3)\,\textrm{nK}=0.21(4) \, T_\mathrm{F}$. The dashed lines correspond to the plane-wave model given by \eq{vZB} at ${\bf q} = m\bf{v}_0$. Conveniently the time origin is shifted to match the phase oscillations with experimental signal. This time shift is justified inasmuch as \eq{vZB} does not incorporate the effect of the laser ramping stage. Its value of about $5\,\mu$s is essentially half the ramping duration. 
{(c)}: Time-of-flight fluorescence images at times $t_1 = 18~\mu$s, $t_2 = 30~\mu$s, $t_3=42~\mu$s, and $t_4=62~\mu$s. These times are indicated by vertical dotted lines in {(b)}.} \label{fig:fig1b} 
\end{figure}

We simulate the Hamiltonian of \eq{general_H} using a degenerate Fermi gas at a temperature $T=30(3)\,$nK, with $T/T_\mathrm{F}=0.21(4)$, where $T_F \approx 143 \,$nK is the Fermi temperature of our gas. After the cooling and preparation sequences, the atoms are in the state $\ket{3}$ \cite{SM}. We switch on the tripod beams to transfer adiabatically all atoms from state $|3\rangle$ to one state in the dark-state manifold. For $v_0=|\mathbf{v}_0|=0$, we expect to populate the dark state $|D_2\rangle$ \cite{SM}. To assess the quality of the adiabatic transfer, we abruptly switch off the tripod beams, let the atoms fall for $9$ ms, record the fluorescence image of the gas that we use for a direct measurement of the velocity distribution. As expected from \eq{D2}, we observe one velocity peak centered at $-2v_r\hat{{\bf e}}_x$ for state $|1\rangle$, a second one at $-v_r(\hat{{\bf e}}_x+\hat{{\bf e}}_y)$ for state $|2\rangle$, and a third one at the origin for state $\ket{3}$; see Fig. \ref{fig:fig1b}(a). By fitting each peak by a Gaussian distribution, we measure the populations $P_a$, which agree at a $98\%$ level with the expected values (1/6,1/6,2/3) inferred from \eq{D2}. We also checked, by adiabatically switching off the tripod lasers, that $95\%$ of the population returns back to state $\ket{3}$. This result indicates a good control of the quantum coherence during the state preparation \cite{SM}.

With the tripod laser detunings, we now fix a certain mean velocity $\mathbf{v}_0$ of the ultracold gas in the moving frame that we characterize by its polar coordinates $(v_0, \theta_0)$ in the tripod laser plane. We let the system evolve in the gauge fields for a time $t$ and measure the bare state populations $P_a(t)$ by the time-of-flight (TOF) technique. The experimentally inferred momentum-averaged velocity in the laboratory frame is simply ${\bf v}_{\mathrm{exp}}(t) = -v_r\left[(2P_1+P_2){\bf \hat{e}}_x+P_2{\bf \hat{e}}_y\right]$. The observed temporal oscillations of the velocity, along the $x$ and $y$ axes, are shown in Fig. \ref{fig:fig1b}(b) for $v_0 = 4\sqrt{2} v_r$ and $\theta_0=0.6\pi$. They constitute the first experimental observation of the noninertial wave packet motion induced by a 2D bulk non-Abelian gauge field on an ultracold gas, without scalar potentials. The damping of the oscillations is due to the finite momentum dispersion $\delta p \sim 0.4 \, \hbar k$ of our degenerate Fermi gas. 
The solid line in Fig. \ref{fig:fig1b}(b) is the theoretical prediction obtained without any fitting parameters by numerically integrating the velocity operator evolutions in the Heisenberg picture, including the finite ramping sequence of the tripod beams, and averaging over the initial momentum distribution \cite{SM}. The dashed lines in Fig.\ref{fig:fig1b}(b) are the theoretical predictions for a wave packet in the gauge fields without any laser ramping stage in the dark state $\ket{D_2}$ and with well-defined momentum $\mathbf{q}=m\mathbf{v}_0$. The mean velocity in the laboratory frame of this plane-wave model reads \cite{cserti2006unified}
\begin{equation} \label{eq:vZB}
{\bf v}(t)  = v_r \, {\bf u}_1(\theta_0) + v_r f(\theta_0) \cos\omega t \, \hat{{\bf e}}_{\theta_0},
\end{equation} 
where the last term captures the noninertial effect with
 \begin{equation} \label{eq:f}
       	f(\theta_0) = \frac{\cos\theta_0 - \sin\theta_0}{2(2 + \cos 2\theta_0)},
\end{equation}	
and
\begin{equation} \label{eq:omega}
   \omega = \frac{2kv_0}{3} \, \sqrt{2+\cos2\theta_0}. 
\end{equation}
The complete derivation of \eqr{vZB}{omega} and the expression for ${\bf u}_1(\theta_0)$ can be found in Ref. \cite{SM}. As shown in Fig.~\ref{fig:fig1b}(b), when the boost momentum $mv_0$ is large compared to the momentum dispersion of the gas, $\delta p\ll mv_0$, the oscillatory motion at short time is well captured by the plane-wave model, whereas its amplitude is slightly overestimated because of the finite ramping time in the experiment \cite{SM}. We will now confront our experimental data to the plane-wave model only.

\begin{figure}
\includegraphics[width=0.47\textwidth]{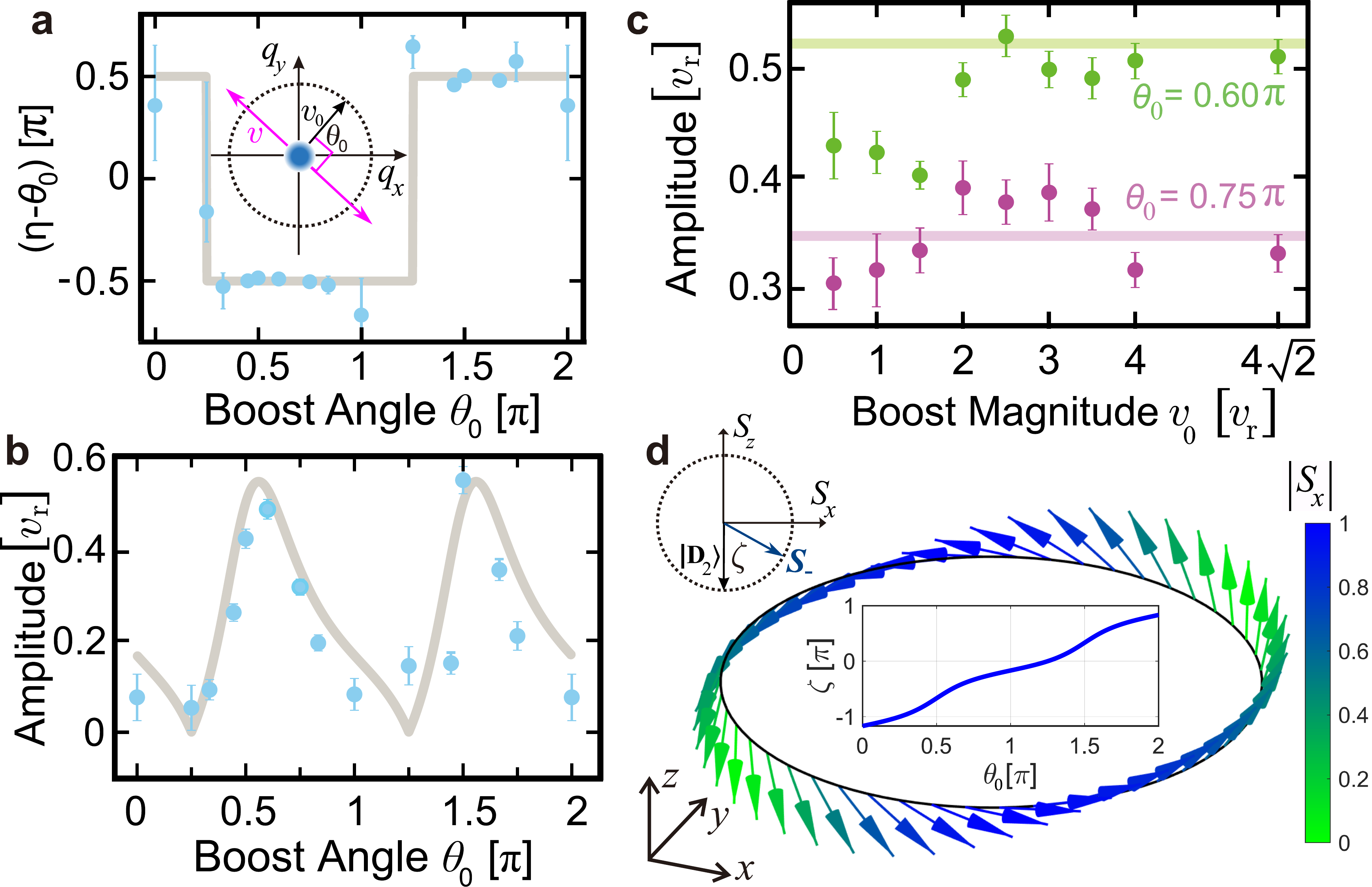}
\caption{{Direction and amplitude of the wave packet oscillation and spin texture.}  
{ (a)} The oscillatory velocity direction $\eta$ as function of $\theta_0$. The blue points are the experimental data whereas the grey plain curve represents the plane-wave model prediction. The inset shows the direction locking of the oscillation at $\eta=\theta_0\pm\pi/2$.
{(b)}: Velocity oscillation amplitude as a function of the boost velocity polar angle $\theta_0$ for $v_0= 4\sqrt{2} v_r$ (points). The solid gray line is the plane-wave prediction $|f(\theta_0)|$ of \eq{f}. 
{ (c)} Velocity oscillation amplitude as a function of $v_0$ at angles $\theta_0=0.6\pi$ (green points) and $\theta_0=0.75\pi$ (magenta points). The green and magenta solid lines are the theoretical predictions from the plane-wave model. The data points in {(a)}, { (b)}, and {(c)} are obtained with a damped-sinusoidal fit function representing the time evolution of the mean velocity of the atoms [see example in Fig. \ref{fig:fig1b}(b)]. All error bars represent 1 standard deviation of uncertainty.
{ (d)}  Spin texture ${\bf S}$ of the lower energy branch (arrows), \eq{texture}, along a circle in the ($Ox,Oy$) momentum plane centered at the Dirac point. The spin orientation lies in the ($Ox,Oz$) plane, and the color code corresponds to the amplitude of the $S_x$ component. The central inset shows the evolution of the angle $\zeta$ of the spin texture in a Bloch sphere representation as a function of $\theta_0$. The angle $\zeta$ is defined with respect to the initial spin orientation $\langle D_2|\mathbf{\boldsymbol{\hat{\sigma}}}|D_2\rangle$; see example on the top-left inset, where $\theta_0=3\pi/2$ and $\zeta=\pi/3$. }\label{fig:fig2}
\end{figure} 

The other central results of this work are the observation of the spin-Hall nature and anisotropy of the 2D motion in momentum space. For this purpose, we vary the mean momentum of the gas in the moving frame via the tripod laser detunings \cite{SM}. We will now discuss these phenomena in detail. At first, we recall that \eq{Evol_V} and \eq{vZB} indicate that the oscillation motion is a manifestation of a spin Hall effect. As such, the velocity oscillation is locked along a direction perpendicular to the momentum $\mathbf{q}$, as it is shown in Fig.~\ref{fig:fig2}(a) for $v_0 = 4\sqrt{2}v_r$. 
Here, for each value of $\theta_0$, we measure the Cartesian coordinates of the oscillating component of the velocity and extract the direction of the motion. We note that the velocity oscillation flips orientation at $\theta_0=\pi/4$ and $\theta_0=5\pi/4$. As we will see below, the amplitude vanishes at these angles. The grey curve is the theoretical prediction from the plane-wave model.

From the Cartesian coordinates of the velocities, we compute the norm and extract the oscillation amplitude as a function of $\theta_0$ that we compare to $v_r|f(\theta_0)|$ as shown in Fig.~\ref{fig:fig2}(b) for $v_0 = 4\sqrt{2}v_r$. Figure~\ref{fig:fig2}(c) shows how the velocity amplitude varies with the boost amplitude $v_0$ for two fixed values of $\theta_0$. When $mv_0$ is no longer significantly larger than $\delta p$, finite momentum dispersion effects kick in and the amplitude departs from the plane-wave model predictions.

To understand the physical origin of the momentum dependence of the velocity oscillations, we derive the local spin textures ${\bf S}_\pm(\theta_0) = \langle \varphi_\pm | \boldsymbol{\hat{\sigma}} |\varphi_\pm\rangle=\mp{\bf S}$, associated to the upper- and lower-energy eigenstates $\ket{\varphi_\pm}$ of the SOC Hamiltonian. 
We have \cite{SM}
\begin{equation} \label{eq:texture}
S_x = \frac{\sqrt{3} \, (\cos\theta_0-\sin\theta_0)}{2\sqrt{2+\cos 2\theta_0}}  \quad S_z = \frac{(3\cos\theta_0+\sin\theta_0)}{2\sqrt{2+\cos 2\theta_0}} 
\end{equation}
and $S_y=0$. In Fig.~\ref{fig:fig2}(d), we show the 3D representation spin texture ${\bf S_-}$ of the lower-energy branch. In the pseudospin language, the initial state $\ket{D_2}$ of our system is the lower spin state. 
For $\theta_0=\pi/4$ and $\theta_0=5\pi/4$, the spin textures are along $Oz$ since $S_x=0$ and the initial state $\ket{D_2}$ identifies with $\ket{\varphi_-}$ and $\ket{\varphi_+}$ respectively [light green arrows in Fig.~\ref{fig:fig2}(d)]. As such, Rabi flopping of the pseudospin cannot occur and the oscillations are suppressed. At these angles, the off-diagonal components of the SOC Hamiltonian 
vanish. In contrast, when $\tan\theta_0 = -3$, so at angles $\theta_0 \approx 0.6\pi$ and $\theta_0 \approx 1.6\pi$, $S_z=0$ and the spin textures are along $Ox$ [blue arrows in Fig.~\ref{fig:fig2}(d)]. At these angles the diagonal terms of the SOC Hamiltonian 
are equal, which corresponds to a resonant excitation in the context of two-level systems. In this case, $\ket{D_2}$ has equal weights on $\ket{\varphi_{\pm}}$ and the oscillation is large though not the largest possible because of the $\theta_0$ dependence in the denominator of $f(\theta_0)$.

\begin{figure}
\includegraphics[width=0.45\textwidth]{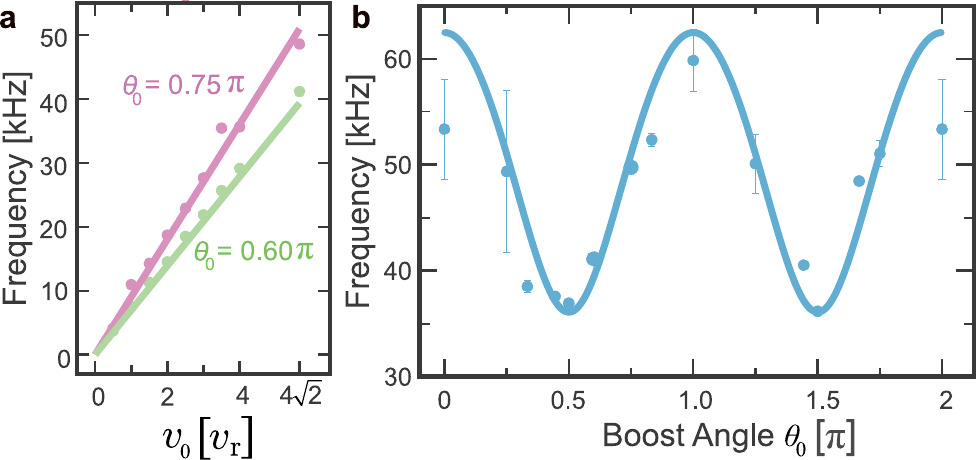}
\caption{{Oscillation frequency.} 
(a) Variation of the oscillation frequency as a function of the magnitude $v_0$ of the boost velocity at fixed angles ($\theta_0=0.60\pi$ for the green data points and $\theta_0=0.75\pi$ for the magenta data points) 
(b) Angular variations of the frequency observed at $v_0 = 4\sqrt{2}v_r$. The solid lines in (b) and (c) are the theoretical predictions inferred from the plane-wave model given by \eq{omega}. The error bars are 1 standard deviation of uncertainty.}
\label{fig:fig3}
\end{figure} 

It is known that a Dirac point is characterized by a winding number that can take two values $\pm1$ \cite{PhysRevLett.121.256402}. From the plane-wave model, this topological number reads $W=(2\pi)^{-1}\oint ({\bf S}\times\nabla_{\theta_0}{\bf S})_yd\theta_0=1$. Seen as a mapping from a circle ($\theta_0$ angle) to another circle (angle $\zeta$ of ${\bf S}$), this reflects the homotopy group of the circle $\pi_1(S^1) = \mathbbm{Z}$ [see insets in Fig.~\ref{fig:fig2}(d)]. It indicates that a spin texture ${\bf S}$ is found along a given direction only twice when $\theta_0$ is circled around $2\pi$. In particular, this given direction can be the initial spin orientation $\langle D_2|\mathbf{\mathbf{\boldsymbol{\hat{\sigma}}}}|D_2\rangle$, explaining why the oscillation amplitude should vanish at least two times along a general loop encircling the Dirac point.

The angular frequency $\omega \propto kv_0$, see \eq{omega} quantifies the energy difference between the upper- and lower-energy branches of the Hamiltonian~\cite{cserti2006unified}. By varying the boost velocity ${\bf v}_0$, the oscillation can be tuned to a suitable frequency scale where it can be easily detected, for example in the kHz range as shown in Fig.~\ref{fig:fig1b}(b). The linear $v_0$ dependence at fixed $\theta_0$, predicted by \eq{omega}, is shown in Fig.~\ref{fig:fig3}(a). Keeping now $v_0$ fixed and circling $\theta_0$ around $2\pi$, the trigonometric variation in \eq{omega} is well reproduced, see Fig.~\ref{fig:fig3}(b).

In conclusion, we have reported on the first experimental observation of a 2D noninertial dynamics in an ultracold atomic gas subject to a non-Abelian $\mathrm{SU}(2)$ gauge field. This result is consistent with predictions of Refs. \cite{zhang2013zitterbewegung,vaishnav2008observing}. We have analyzed in detail the anisotropy of the wave packet motion in momentum space, relating it to a spin-Hall effect. The oscillatory behavior is caused by an interference effect between two spin eigenvalues with different velocities. In a similar way, Schr\"odinger has interpreted the \textit{Zitterbewegung} as interference occurring between the positive an negative energy  of a relativistic particle \cite{schrodinger1930kraftefreie}. In both cases, the oscillation roots in the presence of two noncommutating terms in the Hamiltonian. For the 1D \textit{Zitterbewegung} effect, the mass term does not commute with the spin-orbit component, whereas for the 2D non-Abelian gauge field, the noncommutation occurs between the two spin-orbit components

Our scheme can be extended to $\mathrm{SU}(N)$ systems with $N>2$ \cite{hu2014u}. There, we expect several oscillations frequencies to enter the noninertial dynamics as the different energy branches will not be necessary equally spaced. Very generally, the oscillatory motion would measure the energy differences between these different branches and can develop into a powerful spectroscopic tool to map the energy-branch diagram of such multi-level systems~\cite{valdes2021topological}, in alternative to other existing methods such as rf spectroscopy \cite{cheuk2012spin,huang2016experimental,meng2016experimental}, Bloch oscillation \cite{jotzu2014experimental,li2016bloch}, and Fourier transform spectroscopy \cite{valdes2017fourier,Li1094, Duca288, Tarruell2012}. One could even think of performing selective excitation among energy branches by a clever choice of initial states. 
Finally, the exact nature of noninertial motion in the presence of dynamic gauge fields seems a promising avenue to explore in the future~\cite{Banuls2020,2106.03063}.


M.H. thanks Ulrich Schneider and Bo Song for fruitful discussion. This work was supported by the CQT/MoE funding Grant No. R-710-002-016-271, and the Singapore Ministry of Education Academic Research Fund Tier1 Grant No. MOE2018-T1-001-027 and Tier2 Grant No. MOE-T2EP50220-0008.\\




%

\end{document}


\title{Supplementary Material for ``Wave-packet Dynamics in Synthetic Non-Abelian Gauge Fields"}
\author{Mehedi Hasan}
\affiliation{Nanyang Quantum Hub, School of Physical and Mathematical Sciences, Nanyang Technological University, 21 Nanyang Link, Singapore 637371, Singapore}
\affiliation{MajuLab, International Joint Research Unit IRL 3654, CNRS, Universit\'e C\^ote d'Azur, Sorbonne Universit\'e, National University of Singapore, Nanyang
Technological University, Singapore}
\author{Chetan Sriram Madasu}
\affiliation{Nanyang Quantum Hub, School of Physical and Mathematical Sciences, Nanyang Technological University, 21 Nanyang Link, Singapore 637371, Singapore}
\affiliation{MajuLab, International Joint Research Unit IRL 3654, CNRS, Universit\'e C\^ote d'Azur, Sorbonne Universit\'e, National University of Singapore, Nanyang
Technological University, Singapore}
\author{Ketan D. Rathod}
\affiliation{Centre for Quantum Technologies, National University of Singapore, 117543 Singapore, Singapore}
\affiliation{MajuLab, International Joint Research Unit IRL 3654, CNRS, Universit\'e C\^ote d'Azur, Sorbonne Universit\'e, National University of Singapore, Nanyang
Technological University, Singapore}
\author{Chang Chi Kwong}
\affiliation{Nanyang Quantum Hub, School of Physical and Mathematical Sciences, Nanyang Technological University, 21 Nanyang Link, Singapore 637371, Singapore}
\affiliation{MajuLab, International Joint Research Unit IRL 3654, CNRS, Universit\'e C\^ote d'Azur, Sorbonne Universit\'e, National University of Singapore, Nanyang
Technological University, Singapore}
\author{Christian Miniatura}
\affiliation{MajuLab, International Joint Research Unit IRL 3654, CNRS, Universit\'e C\^ote d'Azur, Sorbonne Universit\'e, National University of Singapore, Nanyang
Technological University, Singapore}
\affiliation{Centre for Quantum Technologies, National University of Singapore, 117543 Singapore, Singapore}
\affiliation{Department of Physics, National University of Singapore, 2 Science Drive 3, Singapore 117542, Singapore}
\affiliation{Nanyang Quantum Hub, School of Physical and Mathematical Sciences, Nanyang Technological University, 21 Nanyang Link, Singapore 637371, Singapore}
\affiliation{Universit\'e C\^ote d'Azur, CNRS, INPHYNI, Nice, France}
\author{Fr\'ed\'eric Chevy}
\affiliation{Laboratoire de Physique de l'\'Ecole normale sup\'erieure, ENS, Universit\'e PSL, CNRS, Sorbonne Universit\'e, Universit\'e de Paris, F-75005 Paris, France}
\author{David Wilkowski}
\email{david.wilkowski@ntu.edu.sg}
\affiliation{Nanyang Quantum Hub, School of Physical and Mathematical Sciences, Nanyang Technological University, 21 Nanyang Link, Singapore 637371, Singapore}
\affiliation{MajuLab, International Joint Research Unit IRL 3654, CNRS, Universit\'e C\^ote d'Azur, Sorbonne Universit\'e, National University of Singapore, Nanyang
Technological University, Singapore}
\affiliation{Centre for Quantum Technologies, National University of Singapore, 117543 Singapore, Singapore}

\begin{abstract}
\end{abstract}

\pacs{}

\keywords{}

\maketitle

\subsection{Experimental methods}

\textbf{Ultracold gas production:} The $^{87}$Sr ultracold gas is produced using standard laser cooling techniques in a magneto-optical trap (MOT) operating on the $^1S_0 \to\,^1P_1$ dipole-allowed transition at $461$ nm (frequency linewidth $32$ MHz), followed by a MOT on the $^1S_0 \to\,^3P_1$ intercombination line at $689$ nm (frequency linewidth $7.5$ kHz), see Refs.~\cite{chaneliere2008three,yang2015high} for more details. Then, the atoms are loaded into a far-off resonant optical dipole trap (ODT) made of two $1064\,$nm laser beams crossing at $70^\circ$ angle. Each beam has a waist of $65\,\mu$m and carry a power of 4~W, leading to trap frequencies around $300\,$Hz and a trap depth about $80~\mu\mathrm{K}$. We load $2.6(1)\times 10^6$ atoms into the ODT at a temperature of $6\,\mu$K. Under a magnetic bias of $23.5\,$G, we perform a partial optical pumping stage in the ground state Zeeman manifold (nuclear spin $I=F=9/2$): Atoms in $m_F>0$ states are transferred into the $m_F = 9/2$ stretched state whereas the remaining atoms in $m_F<0$ states are mostly untouched. A forced evaporative cooling is then implemented during $5.5\,$seconds by exponentially ramping down the power of both beams \cite{desalvo2010degenerate,PhD}. The total number of atoms after evaporation is $N=9.0(5)\times 10^4$ and the gas reaches a final temperature of $T=30(3)~\mathrm{nK}$ with $T/T_\mathrm{F}=0.21(4)$ for the stretched state $m_F = 9/2$. The characterization of the Fermi gas is performed by standard fitting techniques of the momentum distribution with the Thomas-Fermi distribution~\cite{0801.2500}.

\textbf{Fluorescence imaging system:} The velocity distribution of the ultracold gas is probed by fluorescence imaging after a $9\,$ms time-of-flight (TOF) sequence. We turn on, during $10\,\mu$s, an intense counter-propagating pair of resonant $461\,$nm laser beams with orthogonal polarizations. The beams strongly saturate the atomic transition (saturation parameter $s=11$), leading to a fluorescence signal which weakly depends on the laser intensity and on the optical density of the gas. The fluorescence signal is collected on a EMCCD camera through a $2.5\times$ magnification objective with a line of sight normal to the horizontal plane containing the tripod beams. The spatial resolution is  $13\,\mu$m limited by the $16.5\,\mu$m camera pixel size. The velocity resolution of our TOF images is $0.2 v_r$. A signal-to-noise ratio of one on a camera pixel corresponds to a fluorescence signal given by $\sim 70$ atoms.

\textbf{Ground-state population:} After cooling, the atoms are either in the $m_F=9/2$ or in the $m_F<0$ states. The $m_F<0$ atoms are not coupled to the tripod beams and thus remain spectators. However, they induce a bias in the population measurements since the fluorescence technique is not sensitive to the $m_F$ values. To remove this bias, we measure the $m_F<0$ state population by performing a STIRAP stage that transfers the atoms in state $m_F=9/2$ to state $m_F=5/2$ with almost perfect efficiency. During this process, the transferred atoms acquire a velocity kick of $-2v_r$ and can be clearly discriminated in the TOF image from the untouched $m_F<0$ atoms. We found that $51(2)\%$ of the atoms are in $m_F<0$ states. We have removed them in all population measurements pertaining to our gauge fields studies.

When switching on the tripod beams, we want to adiabatically connect the internal state $\ket{m_F=9/2}\equiv\ket{3}$ to a dark state. To do so, we first switch on abruptly the two beams marked 1 and 2 in Fig. 1(b) in the main text. As they are connected to population-empty states, the atomic cloud is unperturbed. Then, we turn on the beam 3 using a linear ramp of $t_{on}=10\,\mu$s, to transfer adiabatically the atoms from state $\ket{3}$ to one state in the dark-state manifold. In the plane-wave approximation and for $v_0=0$, the dark states $\ket{D_1}$ and $\ket{D_2}$ are degenerate (no dynamics). Therefore, we expect to populate the dark state having a non-zero projection on $|3\rangle$, namely $|D_2\rangle$, see \eqs{D1}{D2}. In the experiment, the ultracold gas has a finite momentum distribution with a standard deviation $\delta p \sim 0.4 \hbar k$. However, the SOC characteristic time $\hbar/\delta E_{SOC}\sim\hbar/(v_r\delta p)\sim 40\,\mu$s remains larger than $t_{on}$, so we still expect to populate the dark state $|D_2\rangle$ with high probability~\cite{leroux2018non}. Here, $\delta E_{SOC}$ is the standard deviation of the SOC energy. For quasi-resonant laser beams satisfying $mv_0\gg \delta p$, the SOC characteristic time becomes $\sim\hbar/(mv_rv_0) = (kv_0)^{-1}$ and can be shorter than $t_{on}$. In this situation, the state obtained at $t_{on}$ might rotate in the dark-state manifold and develop a dark state $\ket{D_1}$ component. 
This rotated final state can be tracked with a numerical simulation taking into account the ramp stage as showed in Fig. 1(b) in the main text.

\textbf{Comparison with the plane-wave model:} We confront the experimental data to an analytical plane-wave model assuming an initial internal state $\ket{D_2}$ at momentum $\mathbf{q} = m\bf{v_0}$ in the moving frame (see section D for a complete derivation). This model captures well the dynamics when $mv_0\gg\delta p$, see dashed lines in Fig. 2(b) in the main text. By construction, the model does not include the finite momentum dispersion of the gas and the tripod laser ramping stage. We found that the plane-wave model reproduces well the first oscillations of the experimental signal when shifted by a time $t_{d}\sim 5\,\mu$s which is roughly half the ramp time $t_{on}$. This delay has no consequence on the oscillation frequency estimation but leads to a slight overestimation of its amplitude. Indeed, because of the momentum dispersion of the gas, the oscillation will exponentially damp with a characteristic time $\tau\approx 3/(kv_T\sqrt{5})$ for a Maxwell-Boltzmann distribution with a thermal velocity $v_T=\sqrt{k_BT/m}$ (see Section F). We use an exponentially damped sinusoidal function with a time origin at $t_{on}$ to extract the oscillation amplitude and frequency experimental data. As a consequence, the amplitude shall be compared to $v_r|f(\theta_0)|\exp[-(t_{on}-t_d)/\tau]$ rather than to the plane-wave model prediction $v_r|f(\theta_0)|$. The relative reduction $\exp[-(t_{on}-t_d)/\tau]\sim 0.93$ remains moderate and has been disregarded. We note also that the damping is more pronounced and deviates from an exponential for a Fermi-Dirac distribution (see Section F).

\subsection{Hamiltonian in the dark state manifold}
In the rotating wave approximation (RWA) and in the absence of external potential, the Hamiltonian in space representation of our tripod system reads 
\begin{equation} \label{eq:Htot}
\hat{H}_{tot} = \frac{{\bf \hat{p}}^2}{2m} + \hat{H}_{int} +  \hat{V}_{AL}({\bf r}),
\end{equation}
in the Galilean laboratory frame $S$. Here ${\bf \hat{p}} = -i\hbar \boldsymbol{\nabla}$ and $\hat{H}_{int} = \sum_a \hbar \delta_a \ket{a} \bra{a}$ ($a=1,2,3$), $\delta_a$ being the detuning of the laser addressing the atomic transition $\ket{a} \leftrightarrow \ket{e}$. Assuming that each atomic transition is addressed with the same Rabi coupling strength $\Omega$, the atom-laser interaction term can be written as $\hat{V}_{AL}({\bf r}) = -(\hbar\sqrt{3}\Omega/2) \, \left(\ket{e}\bra{G({\bf r})} + \ket{G({\bf r})}\bra{e}\right)$ where
\begin{equation}
\ket{G({\bf r})} = \frac{1}{\sqrt{3}} \, (e^{-ikx} \ket{1} + e^{-iky} \ket{2} + e^{ikx} \ket{3}).
\end{equation} 
As a consequence, the Hilbert space can be partitioned into a bright manifold $B$, spanned by brights states $\ket{B_\pm({\bf r})} = (\ket{G({\bf r})} \pm \ket{e})/\sqrt{2}$, and an orthogonal dark manifold $D$ which is not coupled to the laser fields. For convenient purposes, we choose the following orthonormal basis in the manifold $D$:
\begin{eqnarray}
\ket{D_1({\bf r})} &=& \frac{1}{\sqrt{2}} \left(e^{-2ikx} \ket{1} - e^{-ik(x+y)} \ket{2} \right), \label{eq:D1}\\
\ket{D_2({\bf r})} &=& \frac{1}{\sqrt{6}} \left(e^{-2ikx} \ket{1} + e^{-ik(x+y)} \ket{2} - 2 \ket{3} \right)  \label{eq:D2}.
\end{eqnarray}
Note that $\{\ket{B_\pm({\bf r})},\ket{D_j({\bf r})}\}$, with ($j = 1,2$), is an orthonormal basis of the Hilbert space. We now use this basis to write $H_{tot}$ in blocks and we assume that our system is initialized in the manifold $D$. By computing the different blocks, it is easy to see that when $|\delta_a| \ll \Omega$ and for sufficiently cold atoms ($k\delta p/m \ll \Omega$), one can use an adiabatic approximation where the coupling to the bright states is suppressed at the time scale explored in the experiment. The ensuing diabatic dynamics is constrained in the dark state manifold and mimics a pseudo-spin $1/2$ evolving under the action of artificial vector and scalar gauge field potentials ${\bf A}_{jk}=i\hbar\bracket{D_j}{\mathbf{\nabla}D_k}$ and $\Phi_{jk}=\left[\hbar^2\bracket{\mathbf{\nabla}D_j}{\mathbf{\nabla}D_k}-({\bf A}^2)_{jk}\right]/2m$ ($j,k = 1,2$)~\cite{dalibard2011colloquium}. The effective Hamiltonian of the system in the dark state manifold then reads 
\begin{equation} \label{eq:GFHamiltonianWithPhi} 
    \hat{H}= \frac{\left({\bf \hat{p}}-\mathbf{\hat{A}}\right)^2}{2m} + \hat{\Phi} + \hat{V}_{\delta},
\end{equation}
where $\hat{V}_{\delta}$ is the matrix restriction of $\hat{H}_{int}$ to the dark state manifold. Using \eqs{D1}{D2}, we find:
\begin{eqnarray}
{\hat{A}_x} &=& \frac{\hbar k}{2} \left[ {\begin{array}{*{20}{c}}
3&{\sqrt{3}/3}\\
{\sqrt{3}/3}&1
\end{array}} \right],\nonumber\\
{\hat{A}_y} &=& \frac{\hbar k}{2} \left[ {\begin{array}{*{20}{c}}
~~~1&{ - \sqrt{3}/3}\\
{ - \sqrt{3}/3}&{~~~1/3}
\end{array}} \right],\nonumber\\
{\hat{V}_\delta} &=& \hbar \left[ {\begin{array}{*{20}{c}}
~~~(\delta_1+\delta_2)/2&{ \sqrt{3} (\delta_1-\delta_2)/6}\\
{ \sqrt{3} (\delta_1-\delta_2)/6}&{~~~(\delta_1+\delta_2+4\delta_3)/6}
\end{array}} \right].
 \label{eq:GaugeFieldComponents0}
\end{eqnarray}
For later purposes, it proves convenient to expand these matrices over unity and the Pauli matrices vector $\hat{\boldsymbol{\sigma}}$: 
\begin{equation} \label{eq:Avec}
{\bf  \hat{A}} = \frac{\hbar k}{3} \begin{pmatrix}
          3 + \sqrt{3} \, \hat{{\bf u}}_x \cdot \hat{\boldsymbol{\sigma}} \\
          1 + \hat{{\bf u}}_y \cdot \hat{\boldsymbol{\sigma}} 
         \end{pmatrix}
\qquad \hat{V}_\delta = \hbar (\Delta + {\bf u}_\delta \cdot \hat{\boldsymbol{\sigma}}), 
\end{equation}
where $\Delta = (\delta_1+\delta_2+\delta_3)/3$ and
\begin{eqnarray} \label{eq:uvec}
\hat{{\bf u}}_x &=& \begin{pmatrix}
          1/2 \\
           0 \\
           \sqrt{3}/2
         \end{pmatrix}
,\quad 
\hat{{\bf u}}_y = \begin{pmatrix}
          -\sqrt{3}/2 \\
           0 \\
           1/2
         \end{pmatrix},\nonumber\\
{\bf u}_\delta &=& \begin{pmatrix}
          \sqrt{3}(\delta_1-\delta_2)/6 \\
           0 \\
           (\delta_1+\delta_2-2\delta_3)/6
         \end{pmatrix},
\end{eqnarray}
in Cartesian coordinates. One may note that the unit vectors $\hat{{\bf u}}_x$ and $\hat{{\bf u}}_y$ are orthogonal.

\subsection{Spin-orbit coupling Hamiltonian in a moving frame}
We now discuss how to make use of the detunings term $\hat{V}_\delta$, to generate the spin-orbit-coupling (SOC) Hamiltonian in Eq.~(6) of the main text. We recall that this Hamiltonian applies in a Galilean frame moving at an arbitrary velocity $-{\bf v}_0$ with respect to the laboratory frame. At first, we show how to create a SOC Hamiltonian in the laboratory frame. Then, we show that a Galilean boost reduces to an extra scalar term that can be compensated for by a proper set of detunings in $\hat{V}_\delta$. Finally we add the previously defined detunings sets to generate the SOC Hamiltonian in Eq.~(6) of the main text.

{\bf Scalar terms compensation:} Expanding the square in \eq{GFHamiltonianWithPhi}, 
\begin{equation}
\hat{H}=\frac{\mathbf{\hat{p}}^2}{2m}-\frac{\mathbf{\hat{p}}\cdot\mathbf{\hat{A}}}{m}+\frac{\mathbf{\hat{A}}^2}{2m}+\hat{\Phi}+\hat{V}_\delta, 
\end{equation}
and noting that our gauge fields satisfy ${\bf \hat{A}}^2/(2m)+\hat{\Phi} = v_r \hat{A}_x$ ($v_r = \hbar k/m$ the recoil velocity), we see that \eq{GFHamiltonianWithPhi} reduces to a SOC Hamiltonian for a set of detunings satisfying $\hat{V}_\delta +v_r \hat{A}_x=0$. Direct inspection using \eq{GaugeFieldComponents0} leads to $\delta_1=-4\omega_r$, $\delta_2 = -2\omega_r$ and $\delta_3 =0$, where $\omega_r=\hbar k^2/(2m)$ is the recoil frequency. We then obtain a SOC Hamiltonian of the form
\begin{equation}
\hat{H}=\frac{\mathbf{\hat{p}}^2}{2m}-\frac{\mathbf{\hat{p}}\cdot\mathbf{\hat{A}}}{m}. 
\end{equation}
We note that a global shift of the detunings $\delta_a \to \delta_a + \delta_0$ adds the constant term $\hbar\delta_0 \openone$ to the Hamiltonian, which does not change the dynamics since an identity term can be removed by redefining the origin of energies. However, to maintain the validity of the adiabatic approximation, the condition $|\delta_a|\ll \omega$ shall be fulfilled. Therefore, this global shift proves useful to minimize $|\delta_a|$ in the experiment and we have implemented it. 

\begin{figure}
\includegraphics[width=0.45\textwidth]{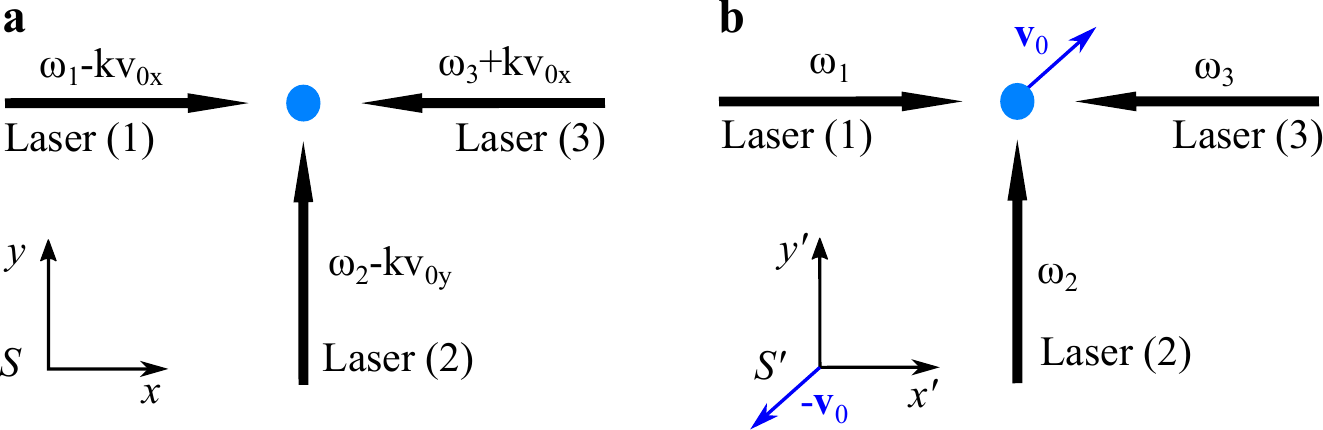}
\caption{ (a) Atom at rest in the laboratory frame $S$ with laser frequencies $\omega_j - {\bf k}_j\cdot{\bf v}_0$. (b) Atom moving at a velocity $\textbf{v}_0$ in the frame $S'$. The tripod beam frequencies are now $\omega_j$.
} \label{fig:figS2}
\end{figure}

{\bf Galilean boost:} Our starting point is again the Hamiltonian in \eq{GFHamiltonianWithPhi}. We perform a Galilean transformation to a new inertial frame $S'$ moving at the constant velocity $-\textbf{v}_0$ with respect to the laboratory frame $S$. It is known \cite{brown1999galilean} that 
the general form of the time-dependent Schr\"{o}dinger's equation in $S$
\begin{equation}
i\hbar\frac{\partial\underline{\Psi}(\mathbf{r},t)}{\partial t}=\left[\frac{(-i\hbar\boldsymbol{\nabla}-\mathbf{\hat{A}})^2}{2m}+V\right]\underline{\Psi}(\mathbf{r},t)
\label{M_Scho_S}
\end{equation}
is preserved under the following transformations:
\begin{eqnarray}
\textbf{r'} &=&\textbf{r}+\textbf{v}_0 t,\quad t' =t,\nonumber\\
\boldsymbol{\nabla}' &=&\boldsymbol{\nabla},\quad \frac{\partial}{\partial t'} =\frac{\partial}{\partial t}-\mathbf{v}_0\cdot\boldsymbol{\nabla},\nonumber\\
\mathbf{\hat{A}}' &=&\mathbf{\hat{A}},\quad \hat{V}' =\hat{V}+\textbf{v}_0\cdot\mathbf{\hat{A}},\nonumber\\
\underline{\Psi}' &=&\textrm{e}^{i\phi}\underline{\Psi},\quad\textrm{with}\quad \phi = \frac{m}{\hbar}\left(\frac{\mathbf{v}_0^2t}{2}+\mathbf{v}_0\cdot\mathbf{r}\right).
\label{eq:M_Cov_Trans}
\end{eqnarray}
This Galilean covariance corresponds to the $c\rightarrow +\infty$ limit of the Lorentz covariance. The Hamiltonian of  \eq{GFHamiltonianWithPhi} reads in the inertial frame $S'$
\begin{equation} \label{eq:Hprime} 
    \hat{H}' = \frac{\left(\mathbf{\hat{q}}-\mathbf{\hat{A}}\right)^2}{2m} + \hat{\Phi} +\textbf{v}_0\cdot\mathbf{\hat{A}} + \hat{V}_\delta,
\end{equation}
with $\mathbf{\hat{q}} = \mathbf{\hat{p}} + m{\bf v}_0$. Using again $\hat{V}_\delta$, we now compensate the extra scalar term $\textbf{v}_0\cdot\mathbf{\hat{A}}$ in \eq{Hprime} coming from the Galilean Boost. Choosing the following detuning set
\begin{equation}
\delta_1=-k\textbf{v}_0\cdot\hat{{\bf e}}_x,\quad\delta_2=-k\textbf{v}_0\cdot\hat{{\bf e}}_y\quad\delta_3=k\textbf{v}_0\cdot\hat{{\bf e}}_x,
\label{eq:v0}
\end{equation}
we get $\hat{V}_\delta=-\textbf{v}_0\cdot\mathbf{\hat{A}}+\hbar k(\textbf{v}_0\cdot\hat{{\bf e}}_x)\openone$. Here, again, the inessential constant term [$\hbar k(\textbf{v}_0\cdot\hat{{\bf e}}_x)\openone$] can be cast away by a global phase change. As expected, the detunings $\delta_a$ in \eq{v0} correspond to a Doppler shift of $\textbf{k}_j\cdot\textbf{v}_0$ of the tripod beams in the frame lab $S$, where $\textbf{k}_j$ is the wave vector of the tripod laser beam $j$, see Fig.\ref{fig:figS2}.

{\bf Spin-orbit coupling Hamiltonian in the moving frame:} To recover a SOC Hamiltonian 
\begin{equation}\label{eq:Hmov}
\hat{H}' = \frac{\mathbf{\hat{q}}^2}{2m}-\frac{\mathbf{\hat{q}}\cdot\mathbf{\hat{A}}}{m}
\end{equation}
like in Eq.~(6) of the main text, we combine the two transformations above, meaning a compensation of scalar terms $\mathbf{\hat{A}}^2/2m+\hat{\Phi}$ and of the $\textbf{v}_0\cdot\mathbf{\hat{A}}$ term coming from the Galilean boost. We then generate a new $\hat{V}_\delta$ term combining the previously found detunings. The new detunings set reads
\begin{eqnarray}
\delta_1 &=& -2\omega_r-k\textbf{v}_0\cdot\hat{{\bf e}}_x,\nonumber\\
\delta_2 &=& -k\textbf{v}_0\cdot\hat{{\bf e}}_y\nonumber\\
\delta_3 &=& 2\omega_r+k\textbf{v}_0\cdot\hat{{\bf e}}_x,
\end{eqnarray}
up to a global shift of the detunings to maintain the tripod beams as close as possible to resonance. Since the atoms are prepared with a zero average momentum in the laboratory frame, they move with an average momentum $m{\bf v}_0$ in the moving frame, see Fig.\ref{fig:figS2}. This momentum is controlled by the frequencies of the tripod laser beams.

\subsection{Oscillation frequency and spin texture}
Since the gauge fields are space-independent, $[\hat{H}, {\bf \hat{p}}]= [\hat{H}', {\bf \hat{q}}] =0$ and both ${\bf \hat{p}}$ and  ${\bf \hat{q}}$ are conserved. This observation invites to proceed to Fourier space and use a plane wave description of the system obtained here by simply replacing the operators ${\bf \hat{p}}$ and ${\bf \hat{q}}$ by their classical counterparts ${\bf p} $ and ${\bf q}= {\bf p} + m{\bf v}_0$. Using the Pauli decomposition, the SOC Hamiltonian of \eq{Hmov} in Fourier space reads
\begin{equation} \label{eq:Hq}
\hat{H}' \to \hat{H}_{{\bf q}} = E_0({{\bf q}}) - \frac{\hbar \boldsymbol{\omega}({\bf q})}{2} \cdot \boldsymbol{\hat{\sigma}}, 
\end{equation}
where
\begin{eqnarray} \label{eq:Hqdecompo}
E_0({{\bf q}}) &=& \frac{{\bf q}^2}{2m} - \frac{\hbar k}{3m} \, (3q_x+q_y) \nonumber\\
\boldsymbol{\omega} &=& \frac{2k}{3m} \, (\sqrt{3} \, q_x \hat{{\bf u}}_x + q_y \hat{{\bf u}}_y).
\end{eqnarray}
Its eigenvalues are $E_\pm({\bf q}) = E_0({{\bf q}}) \pm \hbar \omega/2$, featuring the angular frequency
\begin{equation}\label{eq:ZBfreq}
\omega = |\boldsymbol{\omega}| = \frac{2kq}{3m} \, \sqrt{2 + \cos 2\theta}, 
\end{equation}
where ($q,\theta$) are the polar coordinates of momentum ${\bf q}$ in the tripod laser plane. The Dirac point of the system, $E_{+} = E_{-}$, is obtained when $\omega =0$, that is at ${\bf q}=0$.

The spin textures of the system are defined by ${\bf S}_\pm = \bra{\varphi_\pm} \boldsymbol{\hat{\sigma}} \ket{\varphi_\pm}$, where $\ket{\varphi_\pm}$ are the eigenvectors of $\hat{H}_{\bf q}$. From $E_\pm = \bra{\varphi_\pm} \hat{H}_{\bf q} \ket{\varphi_\pm} = E_0 - \hbar \boldsymbol{\omega}\cdot {\bf S}_\pm/2$, one readily gets $\boldsymbol{\omega}\cdot {\bf S}_\pm = \mp \omega$ and thus opposite spin textures ${\bf S}_\pm = \mp {\bf S}$  with 
\begin{equation} \label{eq:spintex}
{\bf S}=\boldsymbol{\omega}/\omega = \frac{\sqrt{3} \, \cos\theta \, \hat{{\bf u}}_x + \sin\theta \, \hat{{\bf u}}_y}{\sqrt{2+\cos 2\theta}}.
\end{equation}
From this, Eq.~(10) in the main text easily follows. One may note that ${\bf S}^2=1$.

\subsection{Derivation of the velocity motion} 
In the pseudo-spin description induced by the dark state basis, the initial state of the system is the lower spin state $ \underline{\Psi}_i= \begin{pmatrix} 0 \\ 1 \end{pmatrix}$. Defining $\av{X} = \underline{\Psi}^T_i X \, \underline{\Psi}_i$ where the $T$ superscript means transposition, we have $\mathbf{v}(t)\equiv\av{\mathbf{\hat{v}}(t)}= {\bf q}/m - \av{\mathbf{\hat{A}}(t)}/m$ where $\mathbf{\hat{v}}(t) = ({\bf q} - \mathbf{\hat{A}}(t))/m$ is the velocity operator in the moving frame. Using the Pauli decomposition of $\mathbf{\hat{A}}$, and introducing $\boldsymbol{\Sigma}(t) = \av{\boldsymbol{\hat{\sigma}}(t)}$, we find
\begin{equation} \label{eq:meanA}
\av{\mathbf{\hat{A}}(t)}  = \frac{\hbar k}{3} \begin{pmatrix}
          3 + \sqrt{3} \, \hat{{\bf u}}_x \cdot \boldsymbol{\Sigma}(t) \\
           1 + \hat{{\bf u}}_y \cdot \boldsymbol{\Sigma}(t)
         \end{pmatrix}
\end{equation}
in Cartesian coordinates. Using the Heisenberg picture for the spin operator $\boldsymbol{\hat{\sigma}}(t)$, straightforward algebra leads to the spin precession equation
\begin{equation} \label{eq:SpinPre}
\frac{d\boldsymbol{\hat{\sigma}}(t)}{dt} = \frac{i}{\hbar} [\hat{H}_{\mathbf{\hat{q}}}, \boldsymbol{\hat{\sigma}}] = - \boldsymbol{\omega} \times \boldsymbol{\hat{\sigma}}(t)  
\end{equation}
where we have used the identity $[\hat{\sigma}_a, \hat{\sigma}_b] = 2i\sum_c \epsilon_{abc}\hat{\sigma}_c$ featuring the antisymmetric Levi-Civita symbol $\epsilon_{abc}$.
Trivially, $\boldsymbol{\Sigma}(t)$ obeys the same precession equation. Writing $\boldsymbol{\omega} = \omega \, {\bf S}$, and introducing $\boldsymbol{\Sigma} = \boldsymbol{\Sigma}_{\parallel} + \boldsymbol{\Sigma}_\perp$ with $\boldsymbol{\Sigma}_{\parallel} = (\boldsymbol{\Sigma}\cdot {\bf S}) \, {\bf S}$, the general solution reads
\begin{equation}
\boldsymbol{\Sigma}(t) = \boldsymbol{\Sigma}_{\parallel}(0) + \cos\omega t \, \boldsymbol{\Sigma}_\perp(0) - \sin\omega t \, {\bf S} \times \boldsymbol{\Sigma}_\perp(0)
\end{equation}
With our initial state, we have $\boldsymbol{\Sigma}(0) = - \hat{e}_z$, $ \boldsymbol{\Sigma}_{\parallel}(0)=-S_z {\bf S}$ and $ \boldsymbol{\Sigma}_{\perp}(0)=S_x (\hat{{\bf e}}_y \times {\bf S})$ and ${\bf S} \times \boldsymbol{\Sigma}_\perp(0)=S_x \hat{{\bf e}}_y$. It is easy to check from \eq{meanA} that the time-dependent term is indeed perpendicular to ${\bf q}$, as expected from the discussion around Eq.~(2) in the main text. Furthermore, since $\hat{{\bf u}}_x$ and $\hat{{\bf u}}_y$ have no $y$-component in our case, one also sees from \eq{meanA} that the $\sin\omega t$ term does not contribute to the signal. Tedious, but straightforward, calculations then lead to the velocity in the moving frame
\begin{equation} \label{eq:labvelo}
{\bf v}({\bf q},t) = {\bf q}/m + v_r \, {\bf u}_1(\theta) + v_r f(\theta) \cos\omega t \, \hat{{\bf e}}_\theta
\end{equation}
where
\begin{eqnarray} \label{eq:u1}
{\bf u}_{1x} &=& \frac{\sin 2\theta - \cos 2\theta -5}{4(2 + \cos 2\theta)}, \nonumber\\
{\bf u}_{1y} &=& \frac{3\sin 2\theta - 5\cos 2\theta -7}{12(2 + \cos 2\theta)},\nonumber\\
f(\theta) &=& \frac{\cos\theta - \sin\theta}{2(2 + \cos 2\theta)}.
\end{eqnarray}
To obtain the plane-wave model (${\bf p}=0$), we have ${\bf q} = m{\bf v}_0$, thus $q=mv_0$ and $\theta = \theta_0$, the polar angle of ${\bf v}_0$. For comparison with the experiment, we express the velocity in the laboratory frame replacing ${\bf v}$ by ${\bf v}-{\bf v}_0$, and we recover Eqs.~(7) and (8) of the main text. 

\subsection{Initial momentum distribution of the gas and thermal decoherence} 
Our fermionic strontium gas is prepared in a cross-optical-dipole trap with a mean frequency $\omega_{trap} = 100 \,$Hz. It contains $N= 4.5(2)\times 10^4$ atoms in the $m_F=9/2$ state and its Fermi temperature, defined by $k_BT_F = (6N)^{1/3} \, \hbar\omega_{trap}$, is $T_F \approx 143 \, nK$ ($k_B$ is the Boltzmann constant). Its measured temperature is $T=0.21 \, T_F \approx 30 \, nK$, reaching the quantum degeneracy regime. The Fermi-Dirac momentum distribution in the laboratory frame is given by 
\begin{eqnarray} \label{eq:FD}
F({\bf p},T) &=& \frac{1}{2\pi mk_BT} \, \frac{Li_2(-\zeta E)}{Li_3(-\zeta)},\nonumber\\
E &=& \exp(-\frac{{\bf p}^2}{2mk_BT}),\nonumber\\
Li_3(-\zeta) &=& - \frac{1}{6} \, (T/T_F)^{-3},
\end{eqnarray}
where $\zeta$ is the fugacity ($\zeta \approx 59$ here) and $Li_s(z)$ the $n$-order polylogarithmic function~\cite{0801.2500} satisfying $Li_{s+1} (z)= \int_0^z dt \, Li_s(t)/t$. We have $\int d{\bf p} \, F({\bf p}, T) = 1$. It is easy to see that the momentum variance $\delta p^2 = \int d{\bf p} \, p^2 F({\bf p},T)$ of this distribution is given by
\begin{equation} \label{eq:deltap}
\frac{\delta p^2}{2mk_BT_F} = -6 \, (T/T_F)^4 \int_0^\infty x dx \, Li_2(-\zeta e^{-x}) 
\end{equation}
With our parameters, we find a standard deviation $\delta p \approx 0.4 \, \hbar k$.

To obtain the velocity measured in the lab frame, we simply average \eq{labvelo} over $F({\bf p},T)$. Since, the thermal average momentum is zero, we get 
\begin{eqnarray} \label{eq:thermvelo}
{\bf v}_T(t) &=& \int d{\bf p} \, F({\bf p},T) \, {\bf v}({\bf p},t) \nonumber\\
&=& v_r \int d{\bf p} \, F({\bf p},T) \, [{\bf u}_1(\theta) + f(\theta) \cos\omega t \, \hat{{\bf e}}_\theta]. 
\end{eqnarray} 
Physically, we get the superposition of periodic motions with different oscillation frequencies and different oscillation directions which results in an overall damped motion along some average direction. The dispersion of the oscillation frequencies $\omega$, see \eq{ZBfreq}, is related to the dispersion of momentum ${\bf q} = {\bf p} + m{\bf v}_0$ and is thus scaling like $k\delta p/m$. All in all, we expect a damping time given by
\begin{equation}
\tau_D = \frac{m}{k\delta p} \, h(\theta_0), 
\end{equation}
where the function $h(\theta_0)$ can be computed numerically for a Fermi-Dirac distribution and can be approximated by $3/\sqrt{2}\times\sqrt{(2+\cos 2\theta_0)/(5+4\cos 2\theta_0)}$ for a Maxwell-Bolztmann distribution with same temperature, see Ref. \cite{PhD} for more details. The $\theta_0$-dependence reflects the anisotropy of oscillations in momentum space after thermal average. The number of visible oscillations $\sim \omega \tau_D$, is scaling like $mv_0/\delta p$. As a consequence, when $v_0 \gg \delta p/m \approx 0.4 v_r$, the thermal averaging has little effect on the oscillatory motion and ${\bf v}_T(t)$ is very well approximated by the plane-wave model prediction, Eqs.~(7)-(9) of the main text, where oscillations persist indefinitely.

To obtain the plain curves of Fig.~2b in the main text, we have numerically computed ${\bf v}_T(t)$ with a Monte-Carlo method: Using $T$, $T_F$ and $\zeta$ as inferred from our measurements, we have generated $10^4$ momentum points distributed according to \eq{FD} and computed the sum approximating the integral.

\bibliography{ms2}